\documentclass{article}
\usepackage{epsfig}
\usepackage{amsfonts}
\usepackage{euscript}
\begin{document}

\begin{center}
{\bf
\title
*Two alternative approaches to a stellar interior theory. Which of
them is correct?}
\bigskip

\author "B.V.Vasiliev

\bigskip

\bigskip

{sventa@kafa.crimea.ua}
\end{center}

\bigskip

PACS: 64.30.+i; 95.30.-k
\bigskip
\begin{abstract}
All fundamental physical principles are confirmed by numerous
experiments and practical certainty and  have unambiguous
interpretation. But physics of stars is based on few measured
effects only. It gives some freedom for figments of the
imagination. The goal of this paper is to compare two alternative
astrophysical models with measurement data.
\end{abstract}
\bigskip

\section{Introduction}
All basic problems of physics are thoroughly examined. The total
accuracy of their solutions  is confirmed by numerous direct
experiments and their practical using. Any alterative approach to
these basic problems seems not to be reasonable.  At the first
sight it must be applicable to astrophysics too, because it is a
part of physics. But the situation here ia quite different. It is
accepted in astrophysical community to consider the basis of
modern astrophysics as absolutely reliable and steady. But this
basis was developed in the past, when there were no possibilities
to check them by measuring  stellar parameters and many branches
of physics, like plasma physics, did not exist. It forces to
revisit the basis of stellar physics intently. Nowadays there is
possibility to check basic astrophysical problems by means of
comparison of theoretical models with measurement data. The
astronomers are able  to observe and to measure a few parameters
of stellar radiation and   stellar moving,  which obviously depend
on the state of stellar interior. In the first place there are
parameters  of following phenomena:

(1) the apsidal rotation in binary stars

(2) the spectral dependence of solar seismical oscillations.

From this point of view one can consider also the measurement of
solar neutrino flux as one of similar phenomenon. But its result
can be interpreted ambiguously because there are a bad studied
mechanism of their mutual conversation and it seems prematurely to
use this measurement for a stellar models checking.

\section{Two models of stellar interior}

 It is generally accepted to think that the equilibrium of
substance inside a star can be described by the Euler equation:
\begin{equation}
{\mathbf \nabla} P + \gamma {\mathbf g} = 0,\label{gen1}
\end{equation}
where ${\mathbf g}$ is the gravity acceleration, $P$ is the
pressure and $\gamma$ is the substance density. Here the density
satisfies
\begin{equation}
-4\pi G \gamma=div~\mathbf{g},\label{divg}
\end{equation}
where $G$ is the gravity constant. Based on  this equation
astrophysicists supposed that the pressure and the temperature
inside a star are  growing monotonic with the increase of the
depth. As plasma inside a star can be considered as an ideal gas
at the pressure $P = n~kT$ (n and T are its density and
temperature), it is supposed by all astrophysical models  that the
temperature and the plasma density grow in direction to the star
center. Here the temperature reaches tens millions Kelvin and the
density is approximately hundred times greater than its averaged
value. It is a fundamental statement of modern astrophysics. It is
important to underline that although the matter inside a star can
be described as an ideal gas, it is not a gas. It is
electron-nuclear plasma. Historically the Euler equation
(Eq.({\ref{gen1}})) was formulate and applied to the astrophysical
objects at the time when the term "plasma" did not exist and basic
concepts of plasma physics were not developed. The features of
electron-nuclear plasma can not be described by this equation. The
equilibrium state in general case must take into account  the
balance of all forces applied to the system. As particles of
plasma have masses and electric charges, the gravity action on
plasma can induce its electric polarization $\mathcal{P}$. Taking
into account the gravity induced electric polarization (GIEP), the
equilibrium equation obtains the form:
\begin{equation}
{\mathbf \nabla} P + \gamma {\mathbf g}
+4\pi{\EuScript{P}}~div{\EuScript{P}} = 0\label{gen2}
\end{equation}
Analyzing this equation, it is easy to see that there are at last
two possibilities. At first, the plasma body can exist in self
gravitating field in a state (which can be called Eulerian), when
${\EuScript{P}}=0$ and an equilibrium is determined by
Eq.({\ref{gen1}}). But another equilibrium state is possible when
the gravity force is balanced by the electric force:
\begin{equation}
\gamma {\mathbf g} +4\pi{\EuScript{P}}~div{\EuScript{P}}=
0\label{gen3}
\end{equation}
and the pressure inside the plasma body is constant:
\begin{equation}
{\mathbf \nabla} P = 0\label{p0}.
\end{equation}
(The density of plasma under this condition must be constant too
$\gamma=const$). Since the polarisation which is non-uniformly
distributed in space, can be considered as a distribution of
"bond"charges
\begin{equation}
div\EuScript{P}=\rho_{bond},
\end{equation}
we can rewrite the equilibrium condition in the following way:
\begin{equation}
\gamma {\mathbf g} +\rho_{bond}~E = 0\label{gen31}
\end{equation}
where the intensity of the electric field  can be expressed
through the density of "bond"charge:
\begin{equation}
4\pi \rho_{bond}=div~\mathbf{E}.\label{divgg}
\end{equation}
and we can describe further the equilibrium state by
Eq.({\ref{gen31}}) which looks as more convenient at
consideration.

The theoretical consideration based on the Thomas-Fermi
approximation \cite{V-01} states that the hot dense plasma in a
gravity field must have the equilibrium at the following value of
the bond charge electric density:
\begin{equation}
\rho_{bond}=G^{1/2}\gamma\label{rg}
\end{equation}
and at the electric field intensity
\begin{equation}
{\mathbf E}= \frac{\mathbf g}{G^{1/2}}\label{eg}
\end{equation}

\section{The equilibrium of plasma at GIEP effect}

\subsection{The equilibrium density of hot electron-nuclear plasma}

The self-gravity of a celestial body tends to compress the plasma
in its central region. In this compressing a plasma density
reaches up to $10^{26}$ $particles/cm^3$. Such a density
corresponds to a minimum of the plasma energy \cite{V-01}. At
temperatures about tens million Kelvin, plasma can be considered
as an ideal gas. In accordance with the ideal gas definition,
plasma particles interactions are neglected in this approximation.
To take into account these interactions in next step of
approximation, one can see that two mechanisms of interaction play
the most important role.

1) Electrons are Fermi-particles, and they obey the
Fermi-statistics. They can not occupy levels in the energetic
distribution which another electrons have taken. The value of
correction for this interaction is known, it is given in
Landau-Lifshitz course \cite{LL}. This correction is proportional
to the density of particles in the first power and it is positive,
therefore when one takes it into account, the plasma energy in
this approximation is greater than the ideal gas energy at the
same density and temperature.

2)Electrons and nuclei have uniform space distribution at very
high temperatures (when interactions can be neglected completely).
The same correlation in the space distribution between plasma
particles appears at finite temperatures. It can be described by
introducing a correlation correction which takes into account
electrostatic interaction between nuclei and electrons. This
correction is considered in Landau-Lifshitz course \cite{LL}. This
correction is proportional to particle density in 1/2 and it is
negative, therefore taking it into account leads to the decreasing
of the plasma energy in comparison with the ideal gas energy at
the same density and temperature. If both these corrections are
taken into account, we can see that there is a minimum of plasma
energy at its density \cite{V-01}:
\begin{equation}
\eta=\frac{16}{9\pi}\frac{(Z+1)^3}{r_B^3}\label{eta}
\end{equation}
Where $Z$ is the averaged charge number of nuclei from which
plasma consists, $r_B=\frac{\hbar^2}{m_e e^2}$ is Bohr radius. The
existing of equilibrium plasma density $\eta$ is not caused by
GIEP effect directly. The conclusion about equilibrium density of
hot plasma can be deduced from the standard statistical theory and
has general meaning. Taking into account GIEP effect one concludes
that the plasma density is constant. For equilibrium state this
density is obviously equal to $\eta$ because it corresponds to the
energy minimum.

The equilibrium state with the density $\eta$ is inherent to the
dense plasma at high temperature (about $10^7$ K). This
temperature is characteristic for central region of a star only.
It leads to the conclusion that the core with density $\eta$ is
placed in central region of a star and outside of the core there
is a region where $\nabla P\neq 0$ and the density and the
temperature are change to values which are characteristic for the
star surface.

\subsection{Another
equilibrium parameters of star cores}

 The constancy of the pressure ($\nabla P=0$), which is characteristic for star cores,
needs the constancy of the temperature ($\nabla T= 0$). The value
of the equilibrium temperature can be extracted from  the
temperature dependence of the energy. The temperature dependence
of the  star has two branches with different slopes. At high
temperature, when the energy of radiation  has a important role, a
star as a whole has a positive heat capacity: its energy increases
with increasing of its temperature. At a relatively low
temperature the heat capacity of a star is negative. Here the star
energy decreases while its temperature increases. This behavior of
stellar heat capacity is a well known fact, it is discussed in
Landau-Lifshitz course \cite{LL}. Of course, the own heat capacity
of the star substance in each small volume is positive. One
obtains the negative heat capacity of a star as a whole at taking
into account the gravitational interaction. There is a minimum of
the energy placed between these two branches of the temperature
dependence of the energy with positive and negative slopes. It
determines the equilibrium temperature of the star core
\cite{V-01}:
\begin{equation}
{\mathbb{T}}\approx 2  \cdot 10^7(Z+1)~K.\label{T}
\end{equation}
A similar argumentation gives a possibility to determine the
equilibrium mass and the equilibrium radius of star core
\cite{V-01}.

\section{The apsidal rotation in
binary stars}
\subsection{The rotation of close double stars at Eulerian distribution of
substance}

The apsidal rotation (or periastron rotation) of close binary
stars is a result of their non-Keplerian moving which originates
from the non-spherical form of stars. This non-sphericity has been
produced by a rotation of stars about their axes or by their
mutual tidal effect. The second effect is usually smaller  and can
be neglected.  The first and basic theory of this effect was
developed by A.Cleirault in the beginning of XVIII century. Now
this effect was measured for approximately 50 double stars.
According to Clairault's theory the velocity of periastron
rotation must be  approximately in 100 times faster if a matter is
uniformly distributed inside a star. Reversely, it would be absent
if all star mass is concentrated in a star centrum. To reach an
agreement between the measurement data and calculations, it is
necessary to assume that the density of the substance grows in
direction to the centrum of a star and  here it runs up a value
which is a hundred times greater than a mean density of the star.
Just the same mass concentration of the stellar substance is
supposed by all standard theories of a star interior. It has been
usually considered as a proof of astrophysical models. But it can
be considered as a qualitative argument. To obtain a qualitative
agreement between theory and measurements, it is necessary to fit
parameters of the stellar substance distribution in each case
separately.

\subsection{The apsidal motion of close binary stars at taking into account
the GIEP effect}

In the absence of rotation a star would have a spherical core. The
rotation transforms it in a oblate ellipsoid. Its oblateness can
be calculated \cite{VAP-Per} and  the velocity of periastron
rotation can be obtain according to Clairault formulas:
\begin{equation}
\frac{\mathcal{P}}{\mathcal{U}}\biggl(\frac{\mathcal{P}}{\mathcal{T}}\biggr)^2=\sum_1^2\xi_i,\label{PU},
\end{equation}
where $\mathcal{P}$ is the period of the ellipsoidal rotation of
stars, $\mathcal{U}$ is the period of the periastron rotation. The
parameter $\mathcal{T}$ is the period depending on world constants
only:
\begin{equation}
\mathcal{T}=\sqrt{\frac{243~\pi^3}{80}~\frac{a_B^3}{G~m_p}}\approx
10^4 sec
\end{equation}
and the parameter
\begin{equation} \xi_i=\frac{Z_i}{A_i(Z_i+1)^3}
\end{equation}
depends on chemical composition of star cores. There $Z_i$ and
$A_i$ are the charge and the mass number of nuclei which are
composing the plasma of $i$-star.  Hence the velocity of
periastron rotation depends on the chemical composition of a star
only and it decreases rather sharply with the increasing of Z, so
the periastron rotation of a pair consisting of heavy nuclei must
be indistinguishable. The calculations of $\xi$ for a some light
nuclei are shown in Tab.1.

\bigskip
\begin{center}
    {Table 1.}

\begin{tabular}
{||c|c|c||}\hline\hline
  star1&star2 &$\xi_1+\xi_2$\\
  composed of &composed of&\\\hline
  H & H & .25\\
  H & D & 0.1875\\
  H & He & 0.143\\
  H & hn & 0.125\\
  D & D & 0.125 \\
  D & He & 0.0815 \\
  D & hn & 0.0625 \\
  He & He & 0.037 \\
  He & hn & 0.0185 \\ \hline\hline
\end{tabular}
\end{center}

\begin{figure}
\begin{center}
\includegraphics[scale=0.5]{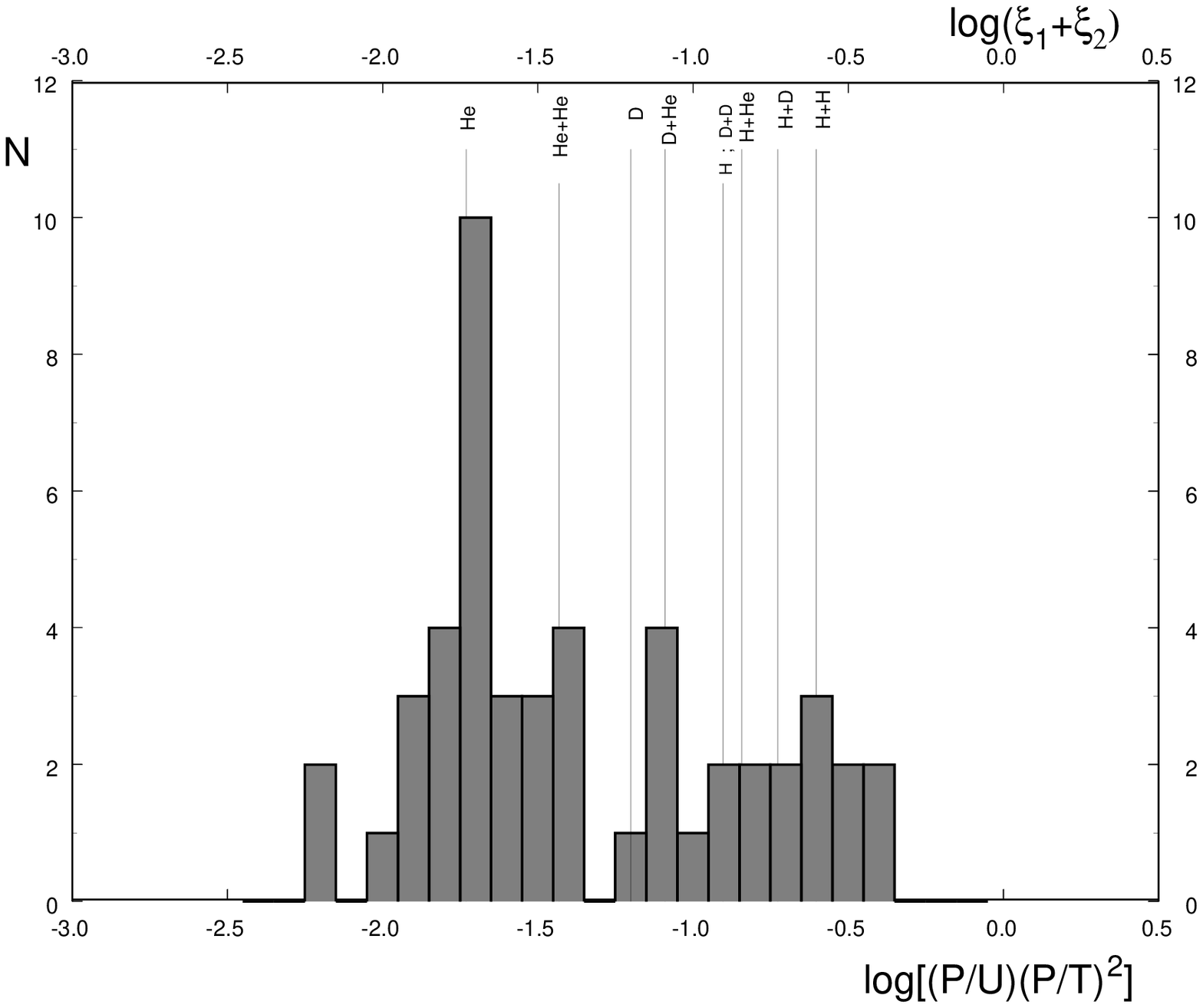}
\caption {The distribution of binary stars on the value of
$(\mathcal{P}/\mathcal{U})(\mathcal{P}/\mathcal{T})^2$.}
\label{periastr}
\end{center}
\end{figure}

 Here  the notation "hn" indicates that the second
component of the couple consists of heavy elements or it is a
dwarf. The distribution of close binary stars on the value of
$(\mathcal{P}/\mathcal{U})(\mathcal{P}/\mathcal{T})^2$ is shown on
Fig.{\ref{periastr}}  in the logarithmic scale.

All these data and references were given to us by
Dr.K.F.Khaliullin (Sternberg Astronomical Institute) and are cited
in \cite{VAP-Per}. On Fig.{\ref{periastr}} the lines mark the
values of parameters (Eq.({\ref{PU}})) for different pairs of
binary stars. It can be seen that the calculated values of the
periastron rotation for stars composed by light elements which is
summarized in Table 1 are in a good agreement with separate peaks
of the measurement data distribution. It is important to underline
that these results were obtained without using of any fitting
parameters (and they are accurate in this regard). Reversely, the
conventional approach to this problem based on Euler equation can
not give any explanation for this measured distribution.

\section{The solar seismical oscillations}
\subsection{ The seismical oscillations at the Eulerian
distribution of a substance}

\begin{figure}
\begin{center}
\includegraphics[scale=0.7]{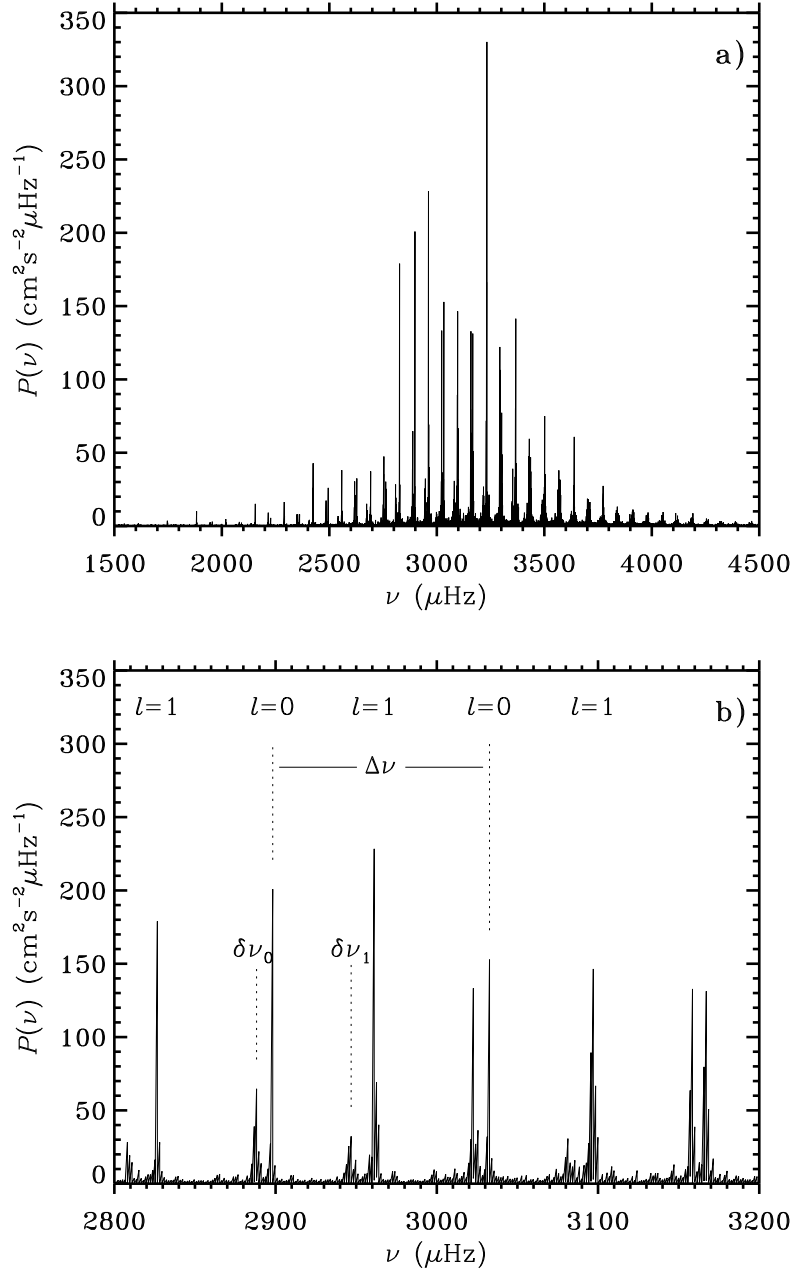}
\caption {$(a)$ The power spectrum of solar oscillation obtained
by means of Doppler velocity measurement in light integrated over
the solar disk. The data were obtained from the BISON network
\cite{bison}. $(b)$ An expanded view of a part of the frequency
range.}\label{bison}
\end{center}
\end{figure}

The measurements \cite{bison} show that the Sun surface is
subjected to a seismic vibration. The most intensive oscillations
have the period about five minutes and the wave length about
$10^4$km or about hundredth part of the Sun radius. Their spectrum
obtained by BISON collaboration is shown on Fig.{\ref{bison}}.


These oscillations are a superposition of a big number of
different modes of resonant acoustic vibrations. It is supposed
that acoustic waves propagate on different trajectories in the
interior of the Sun and have multiple reflection from surface.
With these reflections trajectories of same waves can be closed
and as a result  standing waves are forming.
 Specific features of spherical
body oscillations are described by the expansion in series on
spherical functions. These oscillations can have a different
number of wave lengths on the radius of a sphere ($n$) and  a
different number of wave lengths on its surface which is
determined by the $l$-th spherical harmonic. It is accepted to
describe the sunny surface oscillation spectrum as the expansion
in series \cite{Ch-Del}:
\begin{equation}
\nu_{nlm} \simeq \Delta \nu_0(n+\frac{l}{2}+\epsilon_0)-l(l+1)D_0
+ m\Delta \nu_{rot}.\label{nu}
\end{equation}
 Where the last item is describing
the effect of the Sun rotation and is small. The main contribution
is given by the first item which creates a large splitting in the
spectrum (Fig.{\ref{bison}})
\begin{equation}
\triangle\nu=\nu_{n+1,l}-\nu_{n,l}.
\end{equation}
The small splitting of spectrum (Fig.{\ref{bison}}) depends on the
difference
\begin{equation}
\delta\nu_l=\nu_{n,l}-\nu_{n-1,l+2}\approx (4l+6)D_0.
\end{equation}
A satisfactory agreement of these estimations and measurement data
can be obtained at \cite{Ch-Del}

\begin{equation}
\Delta \nu_0=120~\mu Hz,~ \epsilon_0=1.2,~ D_0=1.5~\mu Hz,~ \Delta
\nu_{rot}=1\mu Hz.\label{del}
\end{equation}

 To obtain these values of parameters $\Delta \nu_0,~ \epsilon_0
è ~D_0$ from theoretical models is not possible. There are a lot
of qualitative and quantitative assumptions used at a model
construction and a direct calculation of spectral frequencies
transforms into a unresolved complicated problem.

\subsection{The oscillation of the solar core at the GIEP effect}
\begin{figure}
\begin{center}
\includegraphics[9cm,22cm][13cm,26cm]{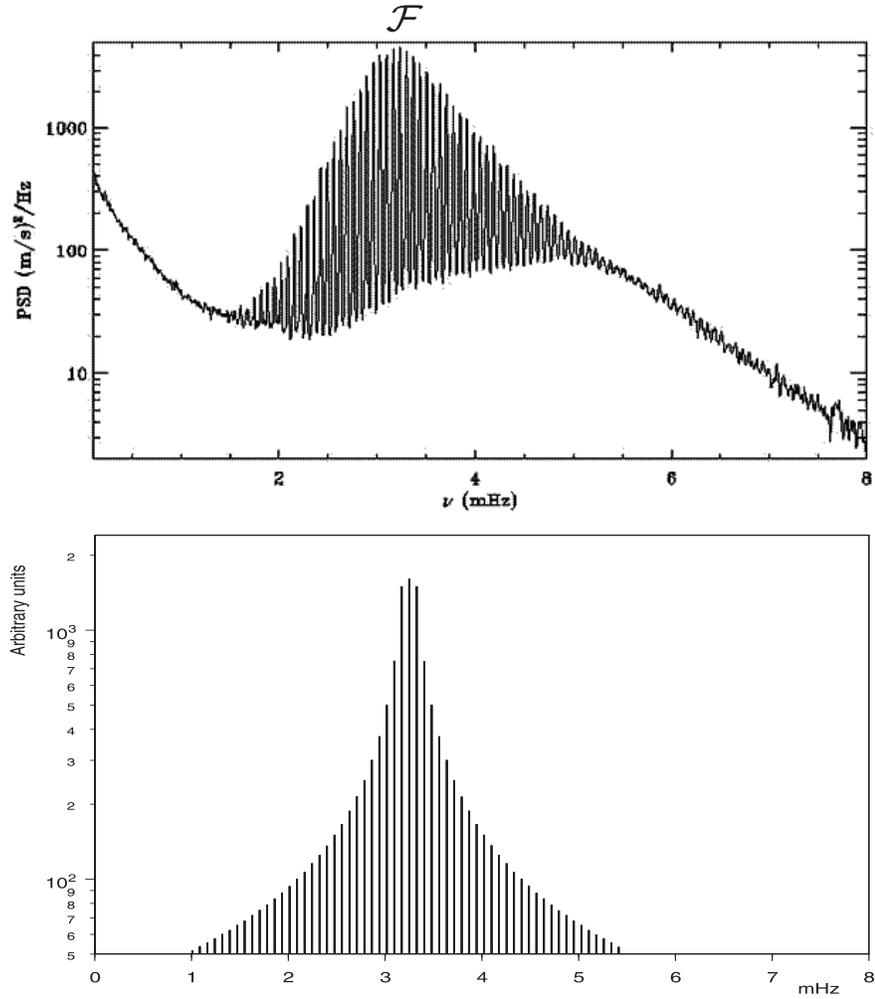}
\vspace{320 pt} \caption{ $(a)$ The spectrum of solar oscillations
obtained by SOHO/GOLF group \cite{GOLF}. $(b)$ The theoretically
obtained spectrum with using Eq.({\ref{ur}}) at $Z=2$ è $A/Z=5$.
}\label{soho}
\end{center}
\end{figure}

Fig.2 shows a central part of the whole spectrum of solar
oscillations which was obtained at a very high frequency
resolution. The whole spectrum of solar oscillations was obtained
at a little worse resolution in the frame of the programm
SOHO/GOLF \cite{GOLF}, and it is shown in Fig.{\ref{soho}}à.

The existence of this spectrum forces to change the view at all
problems of solar oscillations. The theoretical explanation of
this spectrum must give answers as minimum on four questions :

1.Why does the whole spectrum consist from a large number of
equidistant spectral lines?

2.Why does the central frequency of this spectrum ${\cal F}$ is
equal approximately to $\approx 3.23~ mHz$?

3. Why does  this spectrum splitting $f$ is equal approximately to
$67.5~ \mu Hz$?

4. Why does the intensity of spectral lines decrease  from the
central line to the periphery?

The answers to these questions can be obtained from the Sun core
model based on the GIEP effect. According to this model, the Sun
core has the high constant density (Eq.({\ref{eta}})) with radius
${\mathbb{R}}$, mass ${\mathbb{M}}$ at temperature ${\mathbb{T}}$,
which all depend on $Z$ and $A/Z$ only \cite{V-01}.

\subsubsection{ The sound oscillation of the Sun core}

Since the solar core is compressible, the main mode of its
vibration should be elastic sound oscillations of its radius with
the conserved  spherical form of the core at the frequency
$\Omega_s\approx V_{sound}/\mathbb{R}$. The detail calculation
shows that the frequency of this mode \cite{VAP-Sun}
\begin{equation}
\Omega_s =
4.49\biggl\{\frac{10.5}{(3/2)^7\pi}\biggl[\frac{Gm_p}{r_B^3}\biggr]\biggl\langle\frac{A}{Z}\biggr\rangle
\biggl(Z+1\biggr)^3\biggr\}^{1/2}.\label{qb}
\end{equation}
is depending on a chemical composition of the core only. The same
separate frequencies of this mode of the sound radial oscillation
(${\cal F}=\Omega_s/2\pi$) for cores with the different $A/Z$ at
$Z=1$ è $Z=2$ are shown in the third column of Table 2.

\bigskip

\begin{tabular}{||c|c|c||c|c||}\hline\hline
&&${\cal F},mHz$&&${\cal F},mHz$\\
Z&A/Z&&star&\\
& &(calculation Eq.({\ref{qb}}))&&measured\\ \hline
1&1&0.78&$\eta~Bootis$&0.85\\ \hline
&&& The Procion$(A\alpha~CMi)$&1.04\\
1&2&1.10& & \\
&&&$\beta~Hydrae$&1.08\\ \hline 2&2&2.02&&\\\hline
&2.5&2.25&&\\
2&&&$\alpha~Cen~A$&2.37\\\
&3&2.47& & \\\hline
2&3.5&2.67&&\\
2&4&2.85&&\\
2&4.5&3.02&&\\\hline 2&5&3.19&The Sun&3.23\\ \hline\hline
\end{tabular}

{Table 2.}
\bigskip

The measured frequencies of the surface oscillations of separate
stars \cite{Ch-Del} are shown in the right side of Table 2.

Comparing the calculated and measured frequencies, one can
conclude that  the Sun core must be  composed in general by
hellium-10. This conclusion doesn't look so confusing because the
pressure acting in core is running to $10^{18}~{dyn}/{cm^2}$ and
it can induce  the neutronization process in plasma \cite{LL}
which makes neutron-excess nuclei stable. At this chemical
composition we have
\begin{equation}
{\cal F} = \frac{\Omega_s}{2\pi}=3.19~mHz.
\end{equation}
The good agreement with the measurement data gives a possibility
to argue that the central frequency of solar oscillation is
related to the radial oscillations of its core.

\subsubsection{"Phonon-like"low frequency oscillations}
 Another mode of the solar core oscillations is related
to the existence of the equilibrium core density $\eta$
(Eq.{\ref{eta}}). Any deviations of the density from this
equilibrium value, for example which are caused by radial core
oscillations, induce a mechanism of density oscillations around
this equilibrium value with frequency \cite{VAP-Sun}·
\begin{equation}
\omega_\eta=\biggl\{\frac{\sqrt{\pi}~2^4}{\sqrt{10}~(3/2)^7}
\alpha^{3/2}\biggl[\frac{Gm_p}{r_B^3}\biggr]\biggl\langle\frac{A}{Z}\biggr\rangle\biggl[
Z+1\biggr]^{4.5}\bigg\}^{1/2},\label{qm}
\end{equation}
It gives at $Z=2$ and $A/Z = 5$:
\begin{equation}
f_\eta=\frac{\omega_\eta}{2\pi}=66.0~\mu Hz.
\end{equation}
These oscillations of plasma are like phonons in solid bodies. The
excitation of this mechanism can induce oscillations on multiple
frequencies $k\omega_\eta$. Their intensity must be $k$ times
weaker because a population of according levels in energetic
distribution is reversely proportional to their energy
$k\hbar\omega_\eta$. As a result these low frequency oscillations
form spectrum
\begin{equation}
\sum_{k=1} \frac{1}{k}~sin(k\omega_\eta t)
\end{equation}

If these low frequency oscillations $f_\eta$  are induced by the
sound radial oscillation with frequency $\cal{F}$, they will
modulate them. The radial displacements of the solar core surface
form the spectrum
\begin{equation}
u_R\sim \sin~\Omega_s t\cdot\sum_{k=0}
\frac{1}{k}~\sin~k\omega_\eta t~\approx \xi\sin~\Omega_s t +
\sum_{k=1} \frac{1}{k}~\sin~(\Omega_s \pm k \omega_\eta
)t,\label{ur}
\end{equation}
This calculated spectrum is shown on Fig.\ref{soho}b. It is can
been seen from this figure that the model of the Sun based on GIEP
effect gives a possibility to explain all basic details of the
measured spectrum of oscillations and to obtain answers on all
four questions which was formulated above. It is important to
underline that the quantitative agreement between the calculated
spectrum and the measurement data was obtained without using of
any fitting parameters and only taking into account its chemical
composition ($Z=2$ è $A/Z=5$).

\section{ Another measurements which results are depending on a star
interior construction}
\subsection{Stellar masses}
The models of star interior based on the Euler equation does not
give any possibilities to estimate values of star masses. The
theory based on GIEP effect obtains a direct way for the star mass
calculation \cite{V-01}
\begin{equation}
{\mathbb{M}}=1.5^6\biggl(\frac{10}{\pi^{3}}\biggr)^{1/2}
\biggl(\frac{\hbar c}
{Gm_p^2}\biggr)^{3/2} \biggl\langle \frac{Z}{A}\biggr\rangle^2 m_p\nonumber\\
\approx 6.47~ M_{Ch}~
\biggl\langle\frac{Z}{A}\biggr\rangle^2\label{M},
\end{equation}
Where $M_{Ch}=\biggl(\frac{\hbar c}{G
m_p^2}\biggr)^{3/2}m_p=3.71\cdot 10^{33} g$ is the Chandrasekhar
mass.

The frequencies of natural oscillations of the Sun \cite{VAP-Sun}
show that the star core mass is approximately equal to 1/2 of the
whole stellar mass. It gives us a way for the estimation of
stellar masses and for the comparison with measurement results.
The calculated values of star mass depend on one coefficient $A/Z$
only \cite{V-01}.

\begin{figure}
\begin{center}
\includegraphics[scale=0.5]{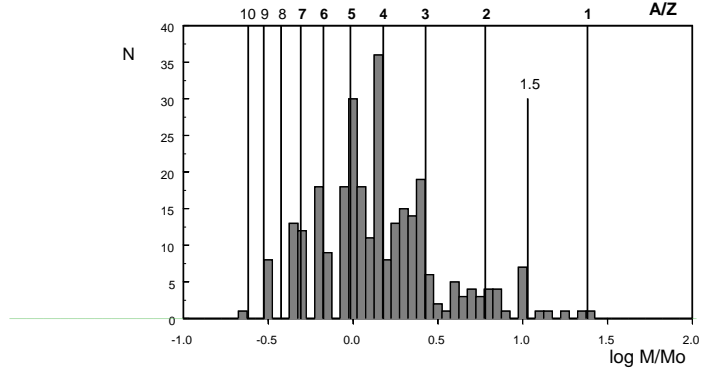}
\caption{The mass distribution of binary stars \cite{Heintz}. On
abscissa, the logarithm of the star mass over the Sun mass is
shown. Solid lines mark masses agree with selected values of $A/Z$
from Eq.({\ref{M}})} {\label{stars}}
\end{center}
\end{figure}

There is no way to determinate the core chemical composition of
far stars, but some predications for it are possible. At first,
there must be no stars which masses exceed the Sun mass more than
one and a half orders, because it accords to limiting mass for
stars consisting from hydrogen with $A/Z = 1$. Secondly, though
the neutronization process makes neutron-excess nuclei stable,
there is no reason to suppose that  stars with $A/Z
> 10$ (and with mass in hundred times less than hydrogen stars) can exist.
Thus, GIEP-theory predicts that the whole mass spectrum must be
placed in the interval from 0.25 up to approximately 25 solar
masses. These predications are verified by measurements quite
exactly. The mass distribution of binary stars is shown on
Fig.{\ref{stars}} \cite{Heintz}. (Using of this data is caused by
the fact that only the measurement of parameters of the binary
star rotation gives the possibility to determine their masses with
satisfactory accuracy). Besides, one can see the presence of
separate pikes for stars with $A/Z = 3; 4; 5...$ and with $A/Z =
3/2$ on Fig.{\ref{stars}}.

It is important to underline that the measured mass of the Sun is
in a good agreement with the claim obtained above  and stating
that the Sun core must be basically composed by hellium-10. This
two measurement - the mass measurement and the frequencies
measurement - build a test practically for the whole GIEP-theory.
The first one tests the formula of mass and the other checks
formulas for the radius of the core and the sound velocity (i.e.
the core density). The agreement of this measurement results
confirms the reliability of the obtained formulas and of the whole
approach.

\subsection{The star magnetic fields}
Another effect which follows from GIEP existence is the generation
of a magnetic field by celestial bodies. These bodies, as they
have electrically polarized cores, must induce magnetic moments
due to their rotation:

\begin{equation}
\mu \approx \frac{Q_{bond}\Omega R^2}{c},
\end{equation}
where $\Omega$ is the rotation velocity,
$Q_{bond}=\frac{4\pi}{3}\rho_{bond} R^3$.

\begin{figure}
\includegraphics[5cm,8cm][9cm,18cm]{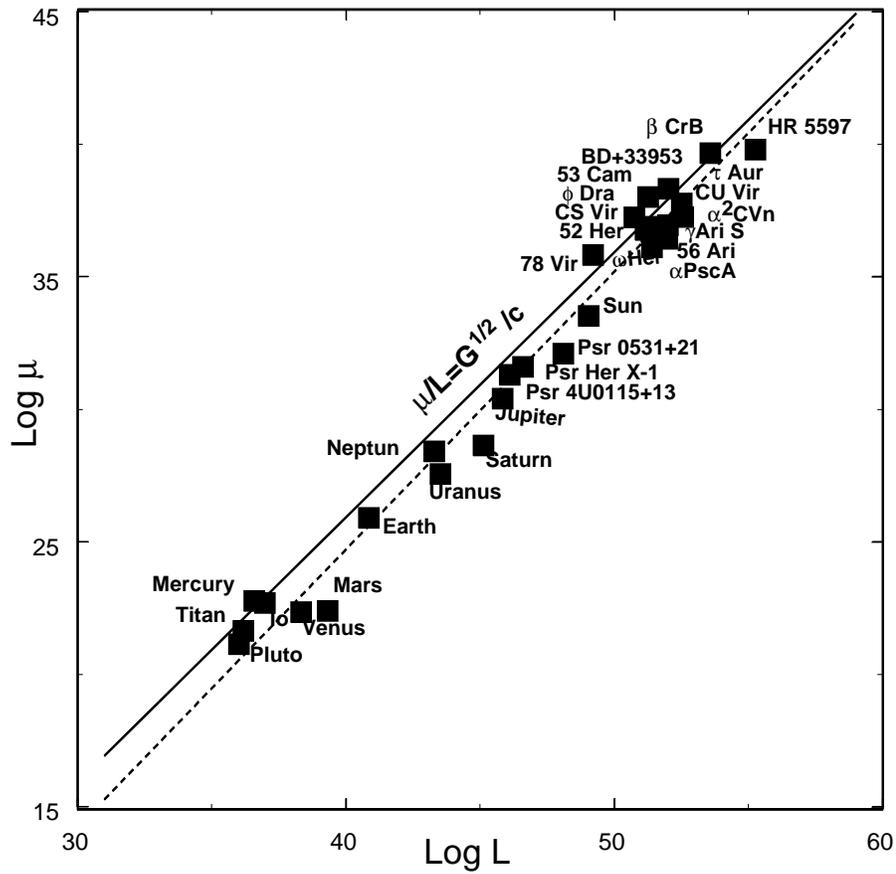}
\vspace{6cm} \caption { The observed values of the magnetic
moments of celestial bodies vs. their angular momenta. On the
ordinate, the logarithm of the magnetic moment over
$Gs\cdot{cm^3}$ is plotted; on the abscissa the logarithm of the
angular momentum over $erg\cdot{s}$ is shown. The solid line
illustrates  Eq.({\ref{3a}}). The dash-dotted line is the fitting
of the observed values. } \label{mL}
\end{figure}

Since the angular momentum of a star

\begin{equation}
L\approx M\Omega R^2,
\end{equation}
we can conclude that the giromagnetic ratios of celestial bodies
must be directly expressed through world constants
\begin{equation}
\vartheta=\frac{\mu}{L}\sim\frac{\sqrt{G}}{c}\label{3a}
\end{equation}

This theoretical predication can be checked by comparison with the
measurement data.  The values of giro-magnetic ratios for all
celestial bodies (for which they are known today) are shown in
Fig.{\ref{mL}}. The data for planets are taken from \cite{Sirag},
the data for stars are taken from \cite{Borra}, and those for
pulsars - from \cite{Beskin}. Therefore, for all celestial bodies
- planets and their satellites, Ap-stars and several pulsars,
which angular momenta distinguish  within more than 20 orders -
the calculated values of the gyromagnetic ratio (Eq.({\ref{3a}}))
agree with measurements quite satisfactorily with the logarithmic
accuracy.

\section{Conclusion}
First conclusion, which we obtain from the above analysis states
that there are four distributions, obtained from measurements,
which depend on properties of the substance inside stars and which
must be explained theoretically. The astrophysical models which
are based on the Euler equation can neither explain the dependence
of the velocity of periastron rotation  from a chemical
composition (Fig.({\ref{periastr}})), nor the star mass
distribution (Fig.({\ref{stars}})), nor their giromagnetic ratios
(Fig.({\ref{mL}})). In this way, the quantitative agreement can be
obtained only by an individual fitting of model parameters. The
explanation of the spectrum of the solar oscillations
(Fig.({\ref{bison}})) by means of series expansion on spherical
harmonics (Eq.({\ref{nu}})) can be considered as a fitting only
because its parameters $\Delta \nu_0,~ \epsilon_0$ and $D_0$ are
free and can not be obtained from the theory.

Quite the contrary, taking into account GIEP effect opens
possibilities for the quantitative explanation with acceptable
accuracy of all measured data without using any fitting
parameters. A good agreement of the relatively simple formulas and
the measurement data has the easy explanation: the cores of stars
consisting of hot dense plasma are well described by the known
ideal gas formulas with small corrections, which are also well
determined by modern plasma physics.

\end{document}